\documentclass[preprint]{iucr}              % DO NOT DELETE THIS LINE
\usepackage{amssymb}
\usepackage[fleqn]{amsmath}
\usepackage{graphicx}
\usepackage{tabularx}
\usepackage{booktabs}
\usepackage{array}
\DeclareMathAlphabet{\mathcalligra}{T1}{calligra}{m}{n}
\def\mathbi#1{\textbf{\em #1}}
\numberwithin{equation}{section}
\def\mathbi#1{\textbf{\em #1}}
\usepackage{color}
\usepackage{ulem}
\usepackage{url}
\usepackage{yfonts}
\newcommand{\SVI}[0]{$\bf{S^{6}}$}
\newcommand{\GVI}[0]{$\bf{G^{6}}$}
\newcommand{\CIII}[0]{$\bf{C^{3}}$}

\newcommand{\Imaginary}[0]{$ \textfrak{I} $}
\newcommand{\Real}[0]{$ \textfrak{R} $}

\journalcode{A}              % Indicate the journal to which submitted
\makeatletter
\font\dummyft@=dummy \relax
\makeatother

\begin{document}                  % DO NOT DELETE THIS LINE

	{\LARGE \emph{\today}} \\
	\title{Delone lattice studies in \CIII, the space of three complex variables}
	%\title{Note on the transformation of three-space basis vectors to  corresponding matrix for Delaunay scalars}
	\shorttitle{properties of C3}
	
	% Authors' names and addresses. Use \cauthor for the main (contact) author.
	% Use \author for all other authors. Use \aff for authors' affiliations.
	% Use lower-case letters in square brackets to link authors to their
	% affiliations; if there is only one affiliation address, remove the [a].

	\cauthor[a]{Lawrence C.}{Andrews}{lawrence.andrews@ronininstitute.org}{}
	\author[b]{Herbert J.}{Bernstein}
	
	\aff[a]{Ronin Institute, 9515 NE 137th St, Kirkland, WA, 98034-1820 \country{USA}}
	\aff[b]{Ronin Institute, c/o NSLS-II, Brookhaven National Laboratory, Upton, NY, 11973 \country{USA}}
	
	% Use \shortauthor to indicate an abbreviated author list for use in
	% running heads (you will need to uncomment it).
	
	\shortauthor{Andrews and Bernstein}
	
	% Use \vita if required to give biographical details (for authors of
	% invited review papers only). Uncomment it.
	
	% lca IUCr id IUCr6401
	%\vita{Author's biography}
	
	% Keywords (required for Journal of Synchrotron Radiation only)
	% Use the \keyword macro for each word or phrase, e.g. 
	% \keyword{X-ray diffraction}\keyword{muscle}
	
	
	% PDB and NDB reference codes for structures referenced in the article and
	% deposited with the Protein Data Bank and Nucleic Acids Database (Acta
	% Crystallographica Section D). Repeat for each separate structure e.g
	% \PDBref[dethiobiotin synthetase]{1byi} \NDBref[d(G$_4$CGC$_4$)]{ad0002}
	
	%\PDBref[optional name]{refcode}
	%\NDBref[optional name]{refcode}
	
	\maketitle                        % DO NOT DELETE THIS LINE
	
	\begin{synopsis}
		The space \CIII{} is explained in more detail than
		in the original description. Boundary transformations
		of the fundamental unit are described in detail. 
		A graphical presentation of the basic coordinates
		is described and illustrated.
	\end{synopsis}
	\newcommand{\si}[0]{$s_1$}
	\newcommand{\sii}[0]{$s_2$}
	\newcommand{\siii}[0]{$s_3$}
	\newcommand{\siv}[0]{$s_4$}
	\newcommand{\sv}[0]{$s_5$}
	\newcommand{\svi}[0]{$s_6$}
	\newcommand{\Svec} [0] {\{\si, \sii, \siii, \siv, \sv, \svi \}}
	\newcommand{\SvecA} [0] {\{-\si, -\si+\sii, \si+\siii, \si+\sv, \si+\siv, \si+\svi \}}
	
	\newcommand{\OPES}[0]{$E^3toS^6$}
	\newcommand{\OPESS}[0]{$$E^3toS^6$$}
	\newcommand{\MSVI}[0]{$M_{S^{6}}$}
	\newcommand{\MEIII}[0]{$M_{E^{3}}$}
	\newcommand{\Plus}[0]{$\textfrak{P}$}	
	\newcommand{\Minus}[0]{$\textfrak{M}$}
	
	\newcommand{\ci}[0]{$c_1$}
	\newcommand{\cii}[0]{$c_2$}
	\newcommand{\ciii}[0]{$c_3$}

	\begin{abstract}
		
		The Delone (Selling) scalars, which are used in 
		unit cell reduction and in lattice type determination,
		are studied in \CIII, the space of three complex variables.
		The three complex coordinate planes are composed of the six
		Delone scalars.
		
		{\bf Note:}  In his later publications, Boris Delaunay used the Russian version of his surname, Delone.\\

	\end{abstract}
	% Appendices appear after the main body of the text. They are prefixed by
	% a single \appendix declaration, and are then structured just like the
	% body text.

	\section{Introduction}
	
	The scalars used by~\citeasnoun{Delaunay1932} in his formulation of Selling reduction ~\cite{Selling1874}
	are (in the conventional order) $b \cdot c$, $a \cdot c$, $a \cdot b$, $a \cdot d$, 
	$b \cdot d$, $c \cdot d$, where $d = -a-b-c$. 
	(As a mnemonic device, 
	observe that the first three terms use
	$\alpha$, $\beta$, and $\gamma$, 
	in that order, 
	and the following terms use $a$, $b$, $c$, in that order.)
	
	~\citeasnoun{andrews2019b} chose to 
	represent the Selling scalars in the space \SVI{},
	\Svec{} (defined in the order above), 
	as a way to create a metric space
	for the measurement of the distance between lattices. 
	~\citeasnoun{andrews2019b} also considered the representation 
	of this space as the
	space of three complex dimensions, \CIII{} or 
	{$\{c_1$}, {$c_2$}, {$c_3\}$}. 	
	
	In \CIII{}, in terms of the Selling scalars, 
	a vector is defined as \{(\si,\siv ), (\sii,\sv),(\siii,\svi)\}, 
	where the real and imaginary parts
	of each are the ``opposite'' scalars 
	according to the definition of~\citeasnoun{Delaunay1932} (see~\citeasnoun{andrews2019a}).
	As a mnemonic device, 
	note that the complex components involve ($\alpha,a$), ($\beta, b$), and ($\gamma,c$).
	Additionally, each complex term uses all 
	four 3-space vectors; for example, $c_1$ is ($b \cdot c$, $a \cdot d$).
	
	~\citeasnoun{andrews2019b} considered the matrix representations 
	of the reflections in \CIII{} (and in \SVI{}). 
	This paper describes the boundary transformations 
	at the edges of the fundamental	unit of \CIII{}. 
	In \SVI{}, the fundamental unit is the all negative orthant, 
	which contains only and	all of the reduced cells. 
	In \SVI{} and \CIII{}, 
	the boundaries located where any \SVI{} scalar 
	(or correspondingly in\CIII{}, the real or imaginary part) 
	equals to zero. The rationale for this work was 
	that it might lend insights into the topology of
	the space of lattices.

	\section{Notation}
	
	Complex numbers will be represented 
	in Cartesian format $(x,y)$, 
	where $x$ is the real part and
	$y$ is the imaginary part.
	
	We will represent a vector 
	in \CIII{} by \{$(x_1,y_1)$, $(x_2,y_2)$, $(x_3, y_3)$\} 
	as an alternative to  \{(\si,\siv ), (\sii,\sv),(\sii,\svi)\}.
	
	Next we define the operators in \CIII{} that 
	will be used in the matrix descriptions of 
	the transformations at the boundaries of the fundamental unit.\\
	See Table ~\ref{tab:operators}.
	
	\begin{table}	
		\label{tab:operators}
		\caption{The Result column values give the products of each operator as applied to $(x_j, y_j)$}.
		\begin{tabular}{c l l l }
			\toprule
			Operator			&Usage						&Result					&Name\\
			\midrule
			$\Minus{}_r $		&  $\Minus{}_r(c_j)$ 		& $(-x_j, -x_j+y_j)$ 	& Minus real \\ 
			$\Minus{}_i $ 		& $\Minus{}_i(c_j)$ 		& $(x_j-y_j,-y_j) $ 	& Minus imag\\ 
			$\textfrak{P}_r$ 	& $ \textfrak{P}_r(c_j)$ 	& $(x_j, x_j)$ 			& Plus real\\ 
			$\textfrak{P}_i$ 	&  $\textfrak{P}_i(c_j)$  	& $(y_j, y_j)$ 			& Plus imag\\ 
			$\Real$ 			&  $\Real(c_j)$ 			& $x_j$ 				& Real \\ 
			$\Imaginary$ 		&  $\Imaginary(c_j)$ 		& $y_j$ 				& Imaginary\\
			\bottomrule
		\end{tabular}		
	\end{table}
	
	\section{Matrices of boundary transformations}
	~~\\
	
	For the boundary at $s_1$: (the real component of $c_1$).
	
	$\begin{bmatrix}
		\Minus{}_r		& 0					& 0 \\
		\Plus{}_r		&  \textrm{i}\Re{}	& \Re{} \\
		\textfrak{P}_r	& \textrm{i}\Im{}	& \Im{}
	\end{bmatrix}$
	$\begin{bmatrix}
		\Minus{}_r	& 0 				& 0 \\
		\Plus{}_r	&\textrm{i}\Im{}	& \Im{} \\
		\Plus{}_r	& \textrm{i}\Re{}	& \Re{} \\
	\end{bmatrix}$ 
	$\begin{bmatrix}
		\label{thirdmatrix}
		\Minus{}_r	& 0			& 0 \\
		\Plus{}_r	&  \Re{}	& \textrm{i}\Re{} \\
		\Plus{}_r	& \Im{}		&\textrm{i}\Im{}
	\end{bmatrix}	$
	$\begin{bmatrix}
		\Minus{}_r	& 0			& 0 \\
		\Plus{}_r	& \Im{}		&\textrm{i}\Im{} \\
		\Plus{}_r	&  \Re{}	&\textrm{i}\Re{} \\
	\end{bmatrix}$ \\
	
	For the boundary at $s_4$: (the imaginary component of $c_1$).
	
	$\begin{bmatrix}
		\Minus{}_i	& 0					& 0 \\
		\Plus{}_i	&  \textrm{i}\Re{}	& \Re{} \\
		\Plus{}_i	& \textrm{i}\Im{}	& \Im{}
	\end{bmatrix}$
	$\begin{bmatrix}
		\Minus{}_i	& 0					& 0 \\
		\Plus{}_i	& \textrm{i}\Im{}	& \Im{} \\
		\Plus{}_i	&  \textrm{i}\Re{}	& \Re{} \\
	\end{bmatrix}$ 
	$\begin{bmatrix}
		\Minus{}_i	& 0			& 0 \\
		\Plus{}_i	&  \Re{}	& \textrm{i}\Re{} \\
		\Plus{}_i	& \Im{}		& \textrm{i}\Im{}
	\end{bmatrix}	$
	$\begin{bmatrix}
		\Minus{}_i	& 0		& 0 \\
		\Plus{}_i	& \Im{}	& \textrm{i}\Im{} \\
		\Plus{}_i	&  \Re{}& \textrm{i}\Re{} \\
	\end{bmatrix}$ \\
	
	For the boundary at \sii{} (the real component of \cii{}):
	
	$\begin{bmatrix}
		
		\textrm{i}\Re{}	& \Plus{}_r		& \Re{} \\
		0				& \Minus{}_r	& 0 \\
		\textrm{i}\Im{}	& \Plus{}_r		& \Im{}
	\end{bmatrix}$
	$\begin{bmatrix}
		\textrm{i}\Im{}	& \Plus{}_r		&  \Im{} \\
		0				& \Minus{}_r	& 0 \\
		\textrm{i}\Re{}	& \Plus{}_r		& \Re{} \\
	\end{bmatrix}$ 
	$\begin{bmatrix}
		\Re{}	&\Plus{}_r		&   \textrm{i}\Re{} \\
		0		& \Minus{}_r	&  0 \\
		\Im{}	&\Plus{}_r		& \textrm{i}\Im{}
	\end{bmatrix}	$ 
	$\begin{bmatrix}
		\Im{}	& \Plus{}_r	& \textrm{i}\Im{} \\
		0		&\Minus{}_r	&  0 \\
		\Re{}	& \Plus{}_r	&  \textrm{i}\Re{} \\
	\end{bmatrix}$ \\

	For the boundary at $s_5$: (the imaginary component of \cii{}).
	
	$\begin{bmatrix}
		\textrm{i}\Re{}	& \Plus{}_i		& \Re{} \\
		0				& \Minus{}_i	& 0 \\
		\textrm{i}\Im{}	& \Plus{}_i		& \Im{}
	\end{bmatrix}$
	$\begin{bmatrix}
		\textrm{i}\Im{}	& \Plus{}_i		&  \Im{} \\
		0				& \Minus{}_i	& 0 \\
		\textrm{i}\Re{}	& \Plus{}_ri	& \Re{} \\
	\end{bmatrix}$ 
	$\begin{bmatrix}
		\Re{}	&\Plus{}_i		&   \textrm{i}\Re{} \\
		0		& \Minus{}_i	&  0 \\
		\Im{}	&\Plus{}_i		& \textrm{i}\Im{}
	\end{bmatrix}	$ 
	$\begin{bmatrix}
		\Im{}	& \Plus{}_i	& \textrm{i}\Im{} \\
		0		&\Minus{}_i	&  0 \\
		\Re{}	& \Plus{}_i	&  \textrm{i}\Re{} \\
	\end{bmatrix}$ \\

	For the boundary at \siii{} (the real component of \ciii{}):
	
	$\begin{bmatrix}
		
		\textrm{i}\Re{}	& \Re{}	& \Plus{}_r \\
		\textrm{i}\Im{}	& \Im{}	& \Plus{}_r  \\
		0				& 0		& \Minus{}_r  \\
	\end{bmatrix}$
	$\begin{bmatrix}
		\textrm{i}\Im{}	&  \Im{}	& \Plus{}_r \\
		\textrm{i}\Re{}	& \Re{}		& \Plus{}_r \\
		0				& 0			& \Minus{}_r \\
	\end{bmatrix}$ 
	$\begin{bmatrix}
		\Re{}	&   \textrm{i}\Re{}	&\Plus{}_r \\
		\Im{}	& \textrm{i}\Im{}	&\Plus{}_r \\
		0 		&  0				& \Minus{}_r \\
	\end{bmatrix}	$ 
	$\begin{bmatrix}
		\Im{}	& \textrm{i}\Im{}	& \Plus{}_r \\
		\Re{}	&  \textrm{i}\Re{}	& \Plus{}_r \\
		0		&  0				&\Minus{}_r \\
	\end{bmatrix}$ \\

	For the boundary at \svi{} (the imaginary component of \ciii{}):
	
	$\begin{bmatrix}
		
		\textrm{i}\Re{}	& \Re{}	& \Plus{}_i \\
		\textrm{i}\Im{}	& \Im{}	& \Plus{}_i \\
		0				& 0		& \Minus{}_i 
	\end{bmatrix}$
	$\begin{bmatrix}
		\textrm{i}\Im{}	&  \Im{}	& \Plus{}_i \\
		\textrm{i}\Re{}	& \Re{}		& \Plus{}_i \\
		0				& 0			& \Minus{}_i \\
	\end{bmatrix}$ 
	$\begin{bmatrix}
		\Re{}	&   \textrm{i}\Re{} &\Plus{}_i \\
		\Im{}	& \textrm{i}\Im{}	&\Plus{}_i \\
		0		&  0				& \Minus{}_i \\
	\end{bmatrix}	$ 
	$\begin{bmatrix}
		\Im{}	& \textrm{i}\Im{}	& \Plus{}_i \\
		\Re{}	&  \textrm{i}\Re{}	& \Plus{}_i \\
		0		&  0				&\Minus{}_i 
	\end{bmatrix}$ \\

	\section{Basics}
	
	The standard representation of the identity operation is
	
	$\bf{c'}$  = 
	$\begin{bmatrix}
		$1$	& $0$	&  $0$ \\
		$0$	& $1$	& $0$ \\
		$0$	& $0$	& $1$ \\
	\end{bmatrix}$ 
	$\bf{c}$. \\

	The identity in \CIII{} can also be written:

	$\bf{c'}$  = 
	$\begin{bmatrix}
		$1$	& $0$						&  $0$ \\
		$0$	& \Re{} +\textrm{i}\Im{}	& $0$ \\
		$0$	& $0$						& \Re{} +\textrm{i}\Im{} \\
	\end{bmatrix}$ 
	$\bf{c}$.
	\\
	
	~\citeasnoun{Delaunay1932} does not 
	consider the boundary transformations 
	in detail. 
	However, he uses them to 
	define the process of Selling reduction. 
	For example in \SVI{}, 
	he lists the following as one of the possible results 
	for a transformation on \si{}:
	\SvecA{}.
	The third boundary transform for $s_1$
	implements this operation and interchanges the
	real part of \ciii{} and the imaginary part of \cii{}:
	
	\begin{center}
		$\begin{bmatrix}
			\Minus{}_r	& 0			& 0 \\
			\Plus{}_r	&  \Re{}	& \textrm{i}\Re{} \\
			\Plus{}_r	& \Im{}		&\textrm{i}\Im{}
		\end{bmatrix}	$
	\end{center}

	Considered in \CIII, Delone's alternate transformation 
	for the \si{} boundary would
	exchange the real of \cii{} 
	with the imaginary part of \ciii{}. That is the fourth
	matrix in the list for \si{}. 
	The other two transformations for \si{} can 
	be generated from the two we have 
	just displayed by the "exchange operation"~\cite{andrews2019b} 
	applied to the second and third \CIII{} coordinates. Delone 
	did not describe the latter two transformations, 
	perhaps because even a single transformation 
	was adequate to implement reduction. He had already listed two.

	\section{Graphical display of projections}
	
	The two-dimensional nature of the three coordinates of \CIII{} 
	suggests their use for graphical display. 
	
	As an example, we use Phospholipase A2 (retrieved from 
	the Protein Data Bank \cite{Bernstein1977}), which has had several similar
	or identical structures determined \cite{LeTrong2007}. \citeasnoun{andrews2019b}
	found additional cases (see Table \ref{PLA2})

\begin{table}
	\begin{tabular}{l c c c c c c c c c}
		\toprule
		PDB id & Centering & a&b&c&$\alpha$&$\beta$&$\gamma$\\
		\midrule
	1DPY & R& 57.98& 57.98& 57.98& 92.02& 92.02& 92.02\\
	1FE5 & R& 57.98& 57.98& 57.98& 92.02& 92.02& 92.02\\
	1G0Z & H& 80.36& 80.36& 99.44& 90&    90&    120\\
	1G2X & C& 80.95& 80.57& 57.1 & 90&    90.35&  90\\
	1U4J & H& 80.36& 80.36& 99.44& 90&    90&    120\\
	2OSN & R& 57.10& 57.10& 57.10& 89.75& 89.75&  89.75\\
	\bottomrule
\end{tabular}	\\
\caption{Phospholipase A2 unit cells}
\label{PLA2}
\end{table}

Below, Figure \ref{fig1} shows the unit cells as 
reported (the centering of lattices has not been removed).
The following figures show various transformations and
embellishments of the reported cells.

\begin{figure}
	\includegraphics[width=1.0\textwidth]{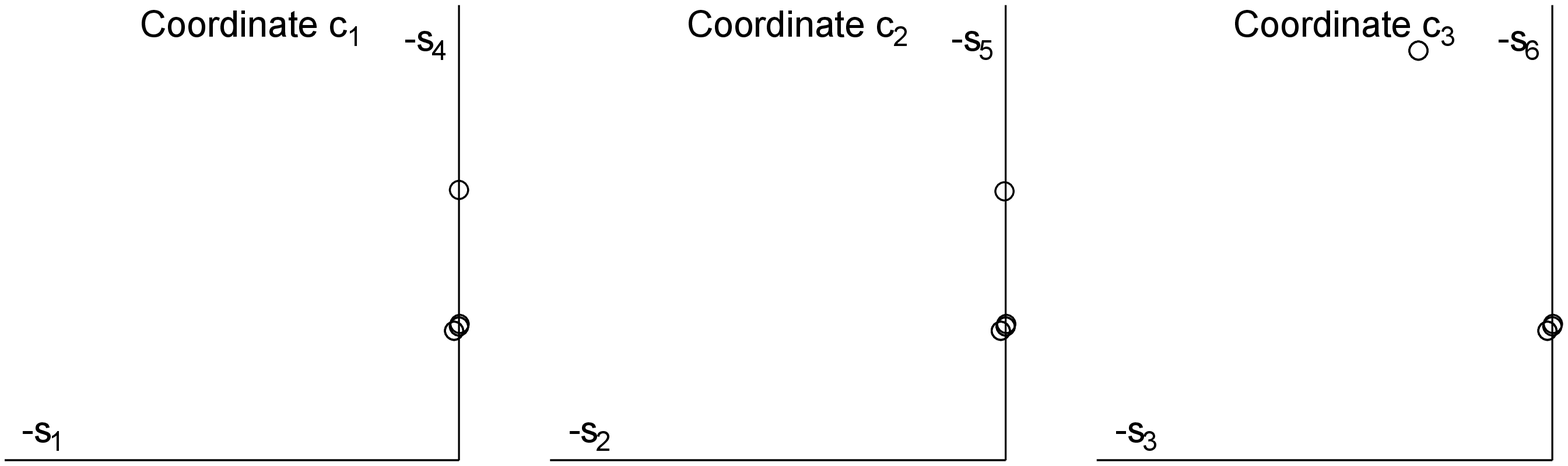}
	\caption{Phospholipase A2 unit cells as reported. }
	\label{fig1}
\end{figure}
	
\begin{figure}
	\includegraphics[width=1.0\textwidth]{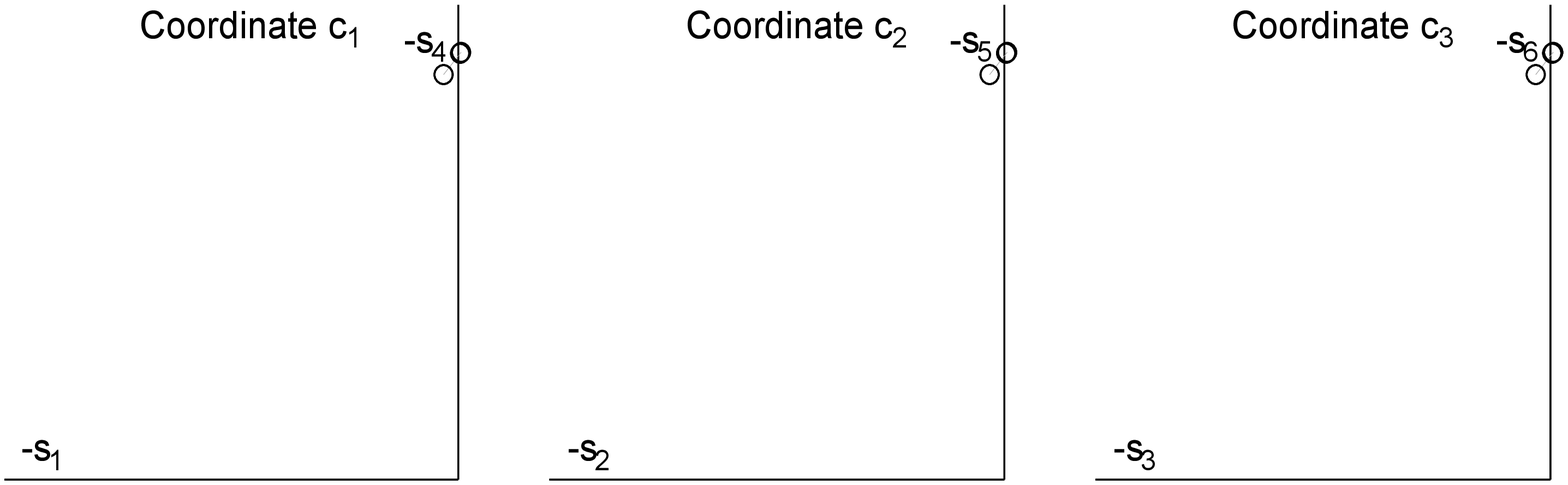}
	\caption{The unit cells Niggli-reduced. The similarity 
	of the 3 projections is indicative of the exact or nearly
exact rhombohedral symmetry.}
	\label{fig2}
\end{figure}
\begin{figure}
	\includegraphics[width=1.0\textwidth]{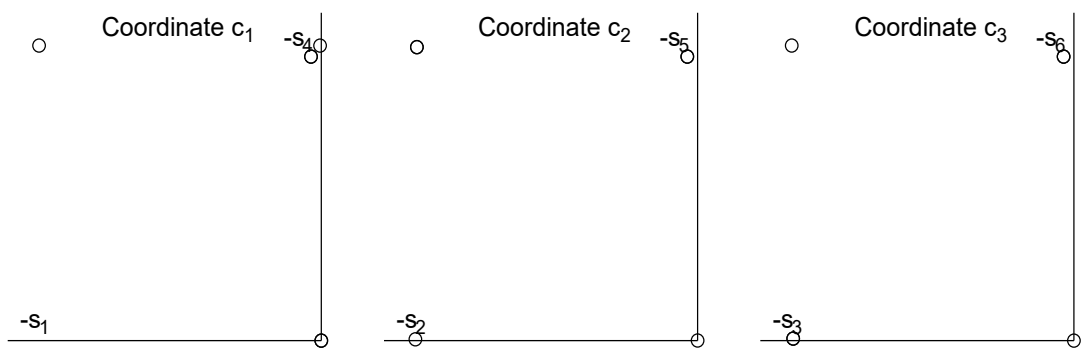}
	\caption{The unit cells Delone reduced. }
	\label{fig3}
\end{figure}

\begin{figure}
	\includegraphics[width=1.0\textwidth]{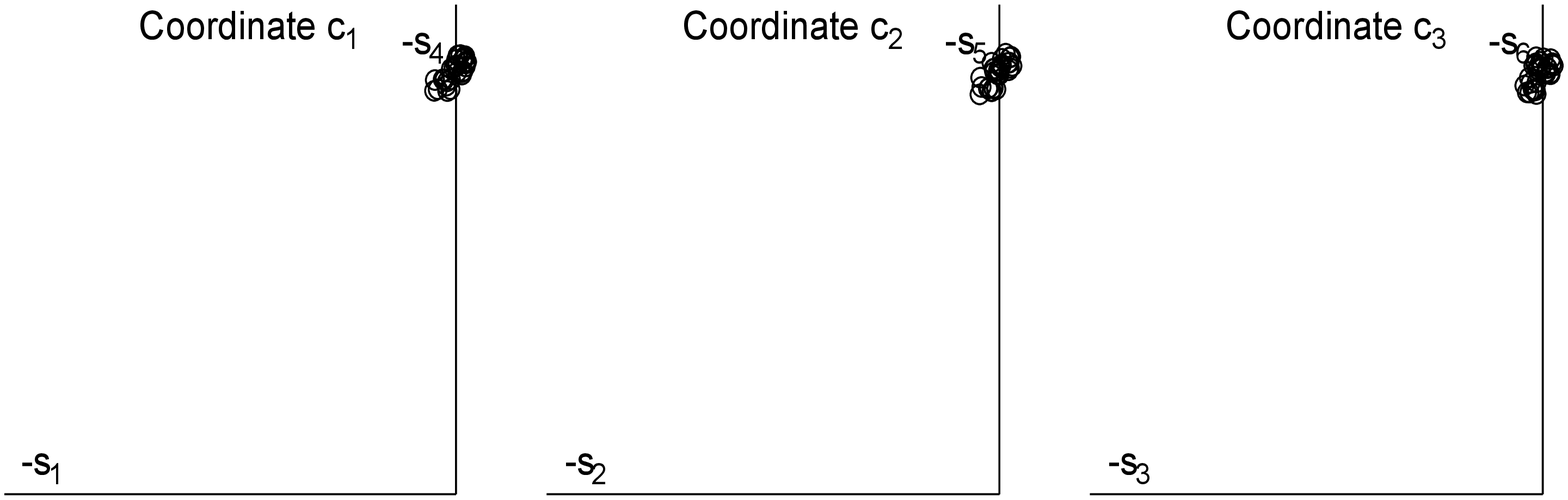}
	\caption{The unit cells, Niggli and five copies were perturbed 2\%
	orthogonally to \SVI{} vector. }
	\label{fig4}
\end{figure}

\begin{figure}
	\includegraphics[width=1.0\textwidth]{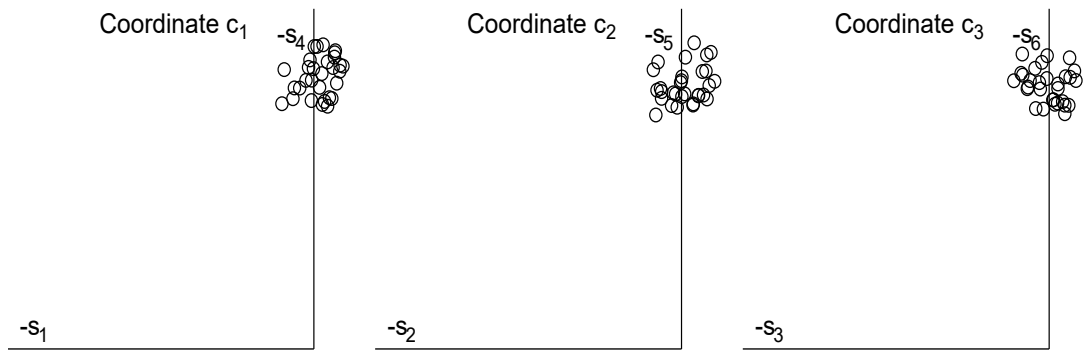}
	\caption{The unit cells Niggli, and five copies were perturbed 10\%
		orthogonally to \SVI{} vector. }
	\label{fig5}
\end{figure}

	\section{Summary}
	
The transformation matrices shown above demonstrate the 
considerable regularity of Selling reduction, as used by 
Delaunay (1932), in comparison to Niggli 
reduction \cite{Niggli1928}. While all the boundaries of the non-positive 
orthant of \SVI{} are essentially the same (and 
related by reflections), the boundaries formed in 
Niggli reduction are of multiple types, and the fundamental unit 
of the space representing Niggli-reduced cells (\GVI{}, see \citeasnoun{Andrews2014}) is non-convex.

The matrices also display one of the unique properties of \CIII{}, 
which makes it a useful conceptual representation. 
For the scalar being transformed, it and its opposite scalar
create a unique row in each transformation matrix. These rows contain only
the minus operator and zeros, highlighting the 
unique relationship of that pair of scalars.

Several aspects of \CIII{} are evident from inspecting the matrices. 
First, each boundary has four possible 
transformations that can be applied. Since each of 
the transformations at boundaries are self-inverse, they are the 
same transformations that would be used in the process of 
cell (lattice) reduction. \citeasnoun{Delaunay1932} and \citeasnoun{Delone1975} give only two choices, presumably for simplicity, 
omitting transformations that use the "exchange" operator 
(see \citeasnoun{andrews2019b}).

	\section{Availability of code}
	
	The $C^{++}$ ~code for \CIII{} is available in github.com, in
	\url{https://github.com/duck10/LatticeRepLib.git}.
	
	%\appendix

	%\section{blah blah blah -- Supplementary Material}
	\ack{{\bf Acknowledgements}}
	
	Careful copy-editing and corrections by Frances C. Bernstein are 
	gratefully acknowledged.
	Our thanks to Jean Jakoncic and Alexei Soares for 
	helpful conversations and access to data and facilities at 
	Brookhaven National Laboratory.
	
	\ack{{\bf Funding information}}      
	
	Funding for this research was provided in part by:  
	US Department of Energy Offices of Biological and 
	Environmental Research and of Basic Energy Sciences 
	(grant No. DE-AC02-98CH10886; grant No. E-SC0012704); 
	U.S. National Institutes of Health (grant No. P41RR012408; 
	grant No. P41GM103473; grant No. P41GM111244; 
	grant No. R01GM117126,
	grant No. 1R21GM129570); Dectris, Ltd.

	\bibliography{Reduced.bib}
	
	\bibliographystyle{iucr}

	%-------------------------------------------------------------------------
	% TABLES AND FIGURES SHOULD BE INSERTED AFTER THE MAIN BODY OF THE TEXT
	%-------------------------------------------------------------------------
	
	% Simple tables should use the tabular environment according to this
	% model
	
	% Postscript figures can be included with multiple figure blocks

\end{document}